\documentstyle[preprint,aps,graphicx]{revtex}

\def\red{
}
\def\black{
}

\def\URLtilde{\lower0.2em\hbox{$\tilde{\phantom{a}}$}}
\def\mycomm#1{\hfill\break\strut\kern-3em{\red\tt ====> #1\black}\hfill\break}
\def\mycommNL#1{\strut\kern-3em{\tt ====> #1}\hfill\break}

\catcode`\@=11 
\def\lsim{\mathrel{\mathpalette\@versim<}}
\def\gsim{\mathrel{\mathpalette\@versim>}}
\def\@versim#1#2{\vcenter{\offinterlineskip
        \ialign{$\m@th#1\hfil##\hfil$\crcr#2\crcr\sim\crcr } }}
\catcode`\@=12 

\def\eqref#1{(\ref{#1})}

\makeatletter
\def\hlinewd#1{\noalign{\ifnum0=`}\fi
\hrule \@height #1 \futurelet \reserved@a\@xhline}
\def\hwhiteline{\noalign
{\ifnum0=`}\fi\hrule
\@height 0pt\vskip 1.0ex\futurelet \reserved@a\@xhline}
\makeatother
\def\gray{\special{ps: 0.40 setgray}}
\def\black{\special{ps: 0.0 setgray}}

\newcommand{\mydraft}{
\newcount\timecount
\newcount\hours \newcount\minutes  \newcount\temp \newcount\pmhours

\hours = \time
\divide\hours by 60
\temp = \hours
\multiply\temp by 60
\minutes = \time
\advance\minutes by -\temp
\def\hour{\the\hours}
\def\minute{\ifnum\minutes<10 0\the\minutes
    \else\the\minutes\fi}
\def\clock{
\ifnum\hours=0 12:\minute\ AM
\else\ifnum\hours<12 \hour:\minute\ AM
\else\ifnum\hours=12 12:\minute\ PM
    \else\ifnum\hours>12
     \pmhours=\hours
     \advance\pmhours by -12
     \the\pmhours:\minute\ PM
     \fi
    \fi
\fi
\fi
}
\def\fullclock{\hour:\minute}
\begin{centering}
\gray
\font\Hugett  =cmtt12 scaled\magstep4
\hbox{\Hugett Draft:\today,\clock}
\black
\end{centering}
\vskip -1.7cm
$\phantom{a}$
} 

\def\beq#1{\begin{equation} \label{#1}}
\def\eeq{\end{equation}}

\newskip\humongous \humongous=0pt plus 1000pt minus 1000pt

\newif\ifdtup

\def\MM{\hbox{${\cal M}^+_{\bar s\kern0.03em\bar s}$}}

\begin{document}
{\tighten
 \preprint {\vbox{
  \hbox{$\phantom{aaa}$}
  \vskip-0.5cm
\hbox{Cavendish-HEP-04/30}
\hbox{TAUP 2788-04}
\hbox{WIS/25/04-SEPT-DPP}
\hbox{ANL-HEP-PR-04-113}
}}

\title{On a possible tetraquark cousin of the $\Theta^+$}

\author{Marek Karliner\,$^{a,b}$\thanks{e-mail: \tt marek@proton.tau.ac.il}
\\
and
\\
Harry J. Lipkin\,$^{b,c}$\thanks{e-mail: \tt
harry.lipkin@weizmann.ac.il} }
\address{ \vbox{\vskip 0.truecm}
$^a\;$Cavendish Laboratory\\
Cambridge University, UK\\
and\\
$^b\;$School of Physics and Astronomy \\
Raymond and Beverly Sackler Faculty of Exact Sciences \\
Tel Aviv University, Tel Aviv, Israel\\
\vbox{\vskip 0.0truecm}
$^c\;$Department of Particle Physics \\
Weizmann Institute of Science, Rehovot 76100, Israel \\
and\\
High Energy Physics Division, Argonne National Laboratory \\
Argonne, IL 60439-4815, USA\\
}
\maketitle
\begin{abstract}%
If a narrow $\Theta^+$ pentaquark exists, it is likely that 
a $ud\bar s$--$ud$
triquark-diquark configuration is a significant component of its wave
function. If so, the mechanism responsible for the binding of a triquark
and a diquark is also likely to bind the triquark to an $\bar s$ antiquark.
We discuss the expected properties of such a $ud\bar s$--$\bar s$
tetraquark meson. In particular, we 
point out that for a $0^+$ isoscalar $ud\bar s\bar s$ meson
 the lowest allowed decay mode is a four-body 
$K K \pi \pi$ channel with a very small phase space and a 
distinctive experimental signature.

\end{abstract} 

\vfill\eject

The recent experimental reports about observation 
\cite{Nakano:2003qx}-\kern-0.5em\cite{H1_Thetac}
and non-observation
\cite{NullRes}
of pentaquarks
have triggered a substantial 
theoretical activity. There have been suggestions as to why some
experiments see the $\Theta^+$ and others don't
\cite{Karliner:2004gr},
but the experimental situation is far from clear and will likely only be
resolved when the results from the new generation of CLAS experiments 
\cite{Rossi:2004rb}
are released.

If a narrow $\Theta^+$ pentaquark does exist, it is likely that
a $ud\bar s$--$ud$ triquark-diquark configuration 
\cite{OlPenta,NewPenta}
is a significant component of its wave
function. If so, the mechanism responsible for the binding of a triquark
and a diquark is also likely to bind the triquark to an $\bar s$ antiquark,
resulting in a manifestly exotic $ud\bar s$--$\bar s$ tetraquark meson 
\MM\,\, with spin zero, strangeness +2 and isospin zero.

     A very simple general argument spelled out in detail below
shows that for any triquark-diquark
model of the $\Theta^+$, the isoscalar $S={+}2$ tetraquark constructed by
replacing the diquark with a strange $\bar s$ antiquark should be above the
$K K$ threshold by 300 MeV more than the $\Theta^+$ is above the $K N$
 threshold.
 
In most theoretical analyses, the $\Theta^+$ pentaquark is assumed to have
a positive parity, corresponding to a triquark and a diquark in a $P$-wave.
If one takes such a configuration and replaces the $ud$ diquark by a
$\bar s$ antiquark, the tetraquark has negative parity. 
It is then easy to see that such a $1^-$
state will have a large decay width to $K^+K^0$.
A $0^-$ or $2^-$ state which cannot decay to $K^+ K^0$ and 
must decay to $K K \pi$ still has a large phase space and is expected to have  
large decay width.

Here we wish to examine the experimental consequences of
another possibility, namely that the  \MM\ tetraquark 
has positive parity. We also briefly discuss the constraints on the
corresponding quark wave function.

A scalar-isoscalar tetraquark with strangeness +2 cannot
decay to anything below $K K \pi\pi$ because of the selection rules that come
from generalized Bose statistics for the $K^+K^0$ system.

      The isoscalar $K^+K^0$ is antisymmetric in flavor and therefore must be
antisymmetric in space. It therefore has odd parity and cannot couple to
an even parity tetraquark. 
The $K K \pi$ state is excluded since any  
$J=0$ state of three pseudoscalar mesons must have odd parity.
A system of three $0^-$ states has odd intrinsic parity. If it is coupled to
$J=0$, it must have even orbital parity because there are only two
independent relative orbital angular momenta and they must be equal to
make $J=0$.

     Thus the lowest $0^+$ state allowed for the decay of an isoscalar 
($ud \bar s \bar s$) tetraquark is a $K K \pi\pi$ state. Moreover, the 
kaons and pions must have a rather nontrivial relative 
angular momentum structure.
If a $KK \pi \pi$ system has isospin zero, the $KK$ and $\pi \pi$
systems must have the same isospin. 
This means they must have opposite parity; if one has
even $L$, the other has odd $L$. Therefore the $KK$ and $\pi \pi$ 
systems must be in a relative $P$-wave to make a $J=0$ state. One possible
channel is a $P$-wave decay
\beq{MtoKkappa}
\MM \,\, \rightarrow\,\,  K^*(1^-) \,\kappa(0^+) \,\, \rightarrow\ K K \pi\pi
\eeq
Thus the lowest decay mode of \MM\ has 
a very distinctive experimental signature.

To make a rough estimate of the \MM\ mass, recall that 
     the difference  $\Delta M(\Theta^+ \rightarrow K^+ n)$
between the $\Theta^+$ mass and the $K^+ n$ mass can be
written as the sum of two terms:
\beq{thetakn}
\Delta M(\Theta^+ \rightarrow K^+ n) = \Delta E(ud \bar s \rightarrow K^+ d) +
\Delta E(dud \rightarrow n)
\eeq
where 
\begin{itemize}
\item[a)]
$\Delta E(ud \bar s \rightarrow K^+ d)$ denotes the energy change due to
splitting 
the triquark into a $K^+$ and a
$d$ quark and moving the $d$ quark next to the color antitriplet diquark.
\item[b)]
$\Delta E(dud \rightarrow n)$ denotes the recombination energy  of the 
color triplet $d$ quark with the
color antitriplet diquark into a neutron.
\end{itemize}

      The difference 
$\Delta M(ud\bar s \bar s \rightarrow K^+ K^0)$ between the $S={+}2$
tetraquark and twice the kaon mass 
can similarly be written as the sum of two analogous terms:
\beq{tetkk}
\Delta M(ud\bar s \bar s \rightarrow K^+ K^0) 
= \Delta E(ud \bar s \rightarrow K^+ d) +
\Delta E(d \bar s \rightarrow K^0)
\eeq
where
\begin{itemize}
\item[a)]
$\Delta E(ud \bar s \rightarrow K^+ d)$ denotes the energy change due to
splitting the triquark into a $K^+$ and a
$d$ quark and moving the $d$ quark next to the color antitriplet diquark.
\item[b)]
$\Delta E(d \bar s \rightarrow K^0)$ denotes the recombination energy  of the 
color triplet $d$ quark with the
color antitriplet antiquark into a kaon.
\end{itemize}

       Although there is no reliable way to estimate the first terms
$\Delta E(ud \bar s \rightarrow K^+ d)$,
it seems reasonable to assume that they are approximately
equal in the two cases. It is the same splitting of the triquark which is
sitting in a color antitriplet color field.
       The second terms have the same color-electric binding of a quark
with an antitriplet. But the hyperfine energy is very different in the two
cases. Combining the d quark with the spin-zero diquark to make a neutron
does not change the hyperfine energy. But combining
the $d$ quark with the strange antiquark to make a kaon gains the kaon hyperfine
energy which is ${3\over4}$ of the $K^*$\raise0.1em\hbox{--}$K$ 
splitting, or about 300 MeV,
\beq{thetatet}
\Delta M(ud\bar s \bar s \rightarrow K^+ K^0) - 
\Delta M(\Theta^+ \rightarrow K^+ n) = 
\Delta E(d \bar s \rightarrow K^0) -
\Delta E(dud \rightarrow n) \approx 300 \,\rm {MeV}
\eeq
This puts the \MM\ above the $K K$ threshold by about 420 MeV, which
gives much more phase space for the decay than for the $\Theta^+$. If the
tetraquark has quantum numbers that forbid $K K$ and allow $K K \pi$, this is
still well above threshold. But if the $\Theta^+$ is a triquark-diquark in an
$S$-wave, this puts the $0^+$ isoscalar \MM\ above the $K K \pi \pi$
threshold by about the same amount that the $\Theta^+$ is above the $K N$
threshold. Moreover, the $K K \pi \pi$ system must contain at least two 
units of angular momentum, coupled to $J=0$. This is likely to make 
\MM\ very narrow.
     Since there has not been any search for this four-body resonance,
it seems reasonable to suggest such a search.

The question of a possible $0^+$ \MM\ state goes back 
to the initial  discussion \cite{OlPenta,NewPenta}
following the  experimental discovery of the $\Theta^+$: the single $S$-wave
cluster is repelled by chromomagnetic effects. Therefore a
diquark-triquark model is chosen to separate the quark pairs of the same flavor. The
$P$-wave gives a centrifugal barrier which helps to keep them apart. But an
$S$-wave is not ruled out with a complicated spatial configuration in a
five-body wave function that keeps them apart.

An example of such complicated spatial correlations arises in the nuclear shell
model, where there is known to be a strong repulsive core in the
nucleon-nucleon interaction. Although the shell-model wave function has
nucleons in relative $S$-states, the effects of this strong repulsion are removed
by methods commonly used in nuclear many-body physics\cite{Bethe} in which the
shell-model wave function is transformed to remove the repulsion.  
Similar arguments can be used to transform the simple S-wave diquark-triquark 
pentaquark wave function and the S-wave antiquark-triquark  tetraquark wave
function to remove the repulsion between identical quark or antiquark pairs.

At this stage we believe that there is little to be gained from doing
a detailed and complicated nuclear physics calculation. But the unusual
experimental signature requiring a four-body resonance is interesting
because it is easy to look for and evidently hasn't been done until now
\cite{ExpNotDone}.

The most favorable initial state for a doubly strange search might be
$K^+p$ which already has one unit of strangeness, since production of a
doubly strange state from an initial state of zero strangeness requires
the creation of two strange $s \bar s$ pairs.  Examples of ``factories"
for inclusively producing doubly strange states are the inclusive
reactions:
\beq{doub1}
K^+\,p \rightarrow \Lambda\, \MM+ X 
\eeq 
\beq{doub2} 
K^+\,p \rightarrow \Sigma \, \MM + X 
\eeq
There is also the exclusive reaction 
\beq{doubex} 
K^+\,p \rightarrow \Sigma^+\,\MM
\eeq
Since $K^+$ and $\MM$ have opposite parity, if the initial state 
in \eqref{doubex} is in an $S$-wave, the final state must be in 
a $P$-wave, etc.

The possibility of  constructing a crypto-exotic tetraquark by replacing the
diquark in a triquark-diquark model of the $\Theta^+$ by a nonstrange antiquark
has been pointed out by  Jaffe\cite{Jafcrypto}. The same simple general
argument used above is even stronger here, since   the relevant threshold is $K
\pi$ and combining the $d$ quark with a nonstrange antiquark to make a pion
gains even more hyperfine energy. This gives an estimate for the cryptoexotic
tetraquark mass as above the $K \pi$ threshold by 400 MeV more than the
$\Theta^+$ is above the $K N$ threshold.

The possibility of constructing an exotic tetraquark by replacing the
diquark in a triquark-diquark model of the $\Theta^+$ by a strange antiquark
has been considered by Close\cite{FrankICHEP}. He does not consider the
scalar tetraquark because of the repulsive short range S-wave interaction. 
Further detailed investigations of exotic tetraquarks by Dudek, Burns and  
Close\cite{Franktet} also do not consider the scalar tetraquark. 

\section*{Acknowledgements}

The research of one of us (M.K.) was supported in part by a grant from the
Israel Science Foundation administered by the Israel
Academy of Sciences and Humanities.
The research of one of us (H.J.L.) was supported in part by the U.S. Department
of Energy, Division of High Energy Physics, Contract W-31-109-ENG-38.

We thank 
Jeff Appel,
Frank Close,
Bob Jaffe,
Sheldon Stone,
Yulia Kalashnikova
and
Bingsong Zou
for discussions about theoretical and experimental aspects of tetraquarks.

%
\catcode`\@=11 
\def\references{
\ifpreprintsty \vskip 10ex
%
\hbox to\hsize{\hss \large \refname \hss }\else
\vskip 24pt \hrule width\hsize \relax \vskip 1.6cm \fi \list
{\@biblabel {\arabic {enumiv}}}
{\labelwidth \WidestRefLabelThusFar \labelsep 4pt \leftmargin \labelwidth
\advance \leftmargin \labelsep \ifdim \baselinestretch pt>1 pt
\parsep 4pt\relax \else \parsep 0pt\relax \fi \itemsep \parsep \usecounter
{enumiv}\let \p@enumiv \@empty \def \theenumiv {\arabic {enumiv}}}
\let \newblock \relax \sloppy
 \clubpenalty 4000\widowpenalty 4000 \sfcode `\.=1000\relax \ifpreprintsty
\else \small \fi}
\catcode`\@=12 


\begin{thebibliography}{99}

\bibitem{Nakano:2003qx}
T.~Nakano {\it et al.}  [LEPS Coll.],
Phys.\ Rev.\ Lett.\  {\bf 91}, 012002 (2003), hep-ex/0301020;

\bibitem{Thetaplus}
V.~V.~Barmin {\it et al.}  [DIANA Coll.],
Phys.\ Atom.\ Nucl.\  {\bf 66}, 1715 (2003)
[Yad.\ Fiz.\  {\bf 66}, 1763 (2003)], hep-ex/0304040;
%
S.~Stepanyan {\it et al.}  [CLAS Coll.],
Phys.\ Rev.\ Lett.\  {\bf 91}, 252001 (2003), hep-ex/0307018;
%
J.~Barth {\it et al.}  [SAPHIR Coll.],
hep-ex/0307083;
%
A.~E.~Asratyan, A.~G.~Dolgolenko and M.~A.~Kubantsev,
hep-ex/0309042;
%
V.~Kubarovsky {\it et al.}  [CLAS Coll.],
[Phys.\ Rev.\ Lett.\  {\bf 92}, 032001 (2004)]
Erratum -- ibid.\  {\bf 92}, 049902 (2004),
hep-ex/0311046;
%
R.~Togoo {\it et al.}, Proc. Mongolian Acad. Sci., {\bf 4} (2003) 2;
%
A.~Airapetian {\it et al.}  [HERMES Coll.],
Phys.\ Lett.\ B {\bf 585}, 213 (2004)
hep-ex/0312044;
%
A.~Aleev {\it et al.}  [SVD Coll.],
hep-ex/0401024;
%
M.~Abdel-Bary {\it et al.}  [COSY-TOF Coll.],
hep-ex/0403011;
%
P.~Z.~Aslanyan, V.~N.~Emelyanenko and G.~G.~Rikhkvitzkaya,
hep-ex/0403044;
%
S.~Chekanov {\it et al.}  [ZEUS Coll.],
hep-ex/0403051;
%
T. Nakano, talk at NSTAR 2004, March 24-27, Grenoble, France,
{\tt http://lpsc.in2p3.fr/congres/nstar2004/talks/nakano.pdf}~;
%
Y.~A.~Troyan {\sl et al.},
hep-ex/0404003.

\bibitem{NA49}
C.~Alt {\it et al.}  [NA49 Collaboration],
hep-ex/0310014.

\bibitem{H1_Thetac}
A.~Aktas {\it et al.}  [H1 Collaboration],
hep-ex/0403017.


\bibitem{NullRes}
J.~Z.~Bai {\it et al.}  [BES Collaboration],
hep-ex/0402012;
%
K.~T.~Knopfle, M.~Zavertyaev and T.~Zivko  [HERA-B Coll.],
hep-ex/0403020 and
I.~Abt {\it et al.}  [HERA-B Collaboration],
hep-ex/0408048;
%
C.~Pinkenburg [for the PHENIX Coll.],
nucl-ex/0404001.
%
P. Hansen [for ALEPH Coll.], talk at DIS 2004,
{\tt http://www.saske.sk/dis04/talks/C/hansen.pdf};
%
T. Wengler [reporting DELPHI Coll. results], talk at
Moriond '04 QCD,
\hfill\break
{\tt http://moriond.in2p3.fr/QCD/2004/WednesdayAfternoon/Wengler.pdf};
%
WA89 Collaboration,
hep-ex/0405042;
%
Y.~M.~Antipov {\it et al.}  [SPHINX Collaboration],
Eur.\ Phys.\ J.\ A {\bf 21}, 455 (2004)
hep-ex/0407026;
%
B.~Aubert {\it et al.}  [BABAR Collaboration],
hep-ex/0408064;
%
M.~J.~Longo {\it et al.}  [HyperCP Collaboration],
hep-ex/0410027;
%
S.~R.~Armstrong,
hep-ex/0410080;
%
K.~Abe  [the Belle Collaboration],
hep-ex/0411005;

\bibitem{Karliner:2004gr}
M.~Karliner and H.~J.~Lipkin,
Phys.\ Lett.\ B {\bf 597}, 309 (2004),
hep-ph/0405002.

\bibitem{Rossi:2004rb}
See e.g.
P.~Rossi  [CLAS Collaboration],
hep-ex/0409057.

\bibitem{OlPenta}
M.~Karliner and H.~J.~Lipkin,
hep-ph/0307243.

\bibitem{NewPenta}
M. Karliner and H.J. Lipkin,
Phys.\ Lett.\ B {\bf 575} (2003) 249.


\bibitem{Bethe}
H. A. Bethe, Phys.\ Rev.\  {\bf 167} (1968) 879.

\bibitem{ExpNotDone}
S. Stone, private communication;
B. Zou, private communication;
J.A. Appel, private communication.

\bibitem{Jafcrypto}
R.~L.~Jaffe, private communication.

\bibitem{FrankICHEP}
F.~E.~Close, talk at ICHEP04.

\bibitem{Franktet}
F.~E.~Close, private communication.

\end{thebibliography}
\end{document}